\documentclass[twocolumn,aps,prb,showpacs]{revtex4}
\usepackage{graphicx}

\begin{document}

\author{Arne Brataas,$^{1}$ Yaroslav Tserkovnyak,$^1$ Gerrit
E. W. Bauer,$^2$ and Bertrand \ I.\ Halperin$^1$}

\affiliation{$^1$Harvard University, Lyman Laboratory of Physics,
Cambridge, Massachusetts 02138} \affiliation{$^2$Department of Applied
Physics and DIMES, Delft University of Technology, 2628 CJ Delft, The
Netherlands}

\title{Spin battery operated by ferromagnetic resonance}

\begin{abstract}
Precessing ferromagnets are predicted to inject a spin current into
adjacent conductors \textit{via} Ohmic contacts, irrespective of a
conductance mismatch with, for example, doped semiconductors. This
opens the way to create a pure spin source (\textquotedblleft spin
battery\textquotedblright) by the ferromagnetic resonance.  We
estimate the spin current and spin bias for different material
combinations.
\end{abstract}

\date{\today}
\maketitle

The research field of magnetoelectronics or spinelectronics strives
to utilize the spin degree of freedom for electronic applications.
\cite{wolf:sci01} Devices made from\ metallic layered systems
displaying the giant \cite{Baibich:prl88} and tunnel magnetoresistance
\cite{Miyazaki:jmmm95} have been proven useful for read-head sensors
and magnetic random-access memories. Integration of such devices with
semiconductor electronics is desirable but difficult because a large
resistivity mismatch between magnetic and normal materials is
detrimental to spin injection.\cite{Schmidt:prb00} Spin injection
into bulk semiconductors has been reported only in optical pump and
probe experiments,\cite{Kikkawa:nat99} and with high-resistance
ferromagnetic injectors \cite {Fieder:nat99} or Schottky/tunnel
barriers.\cite {Monsma:Sc98} In these cases, the injected
spin-polarized carriers are hot and currents are small, however.
Desirable are semiconductor devices with an efficient all-electrical
cold-electron spin injection and detection \textit{via} Ohmic contacts at the
Fermi energy, just as has been realized by Jedema \textit{et al.} for
metallic devices.\cite{Jedema:nat01}

We introduce a concept for DC spin-current injection into arbitrary
conductors through Ohmic contacts, which does not involve net charge
currents. The spin source is a ferromagnetic reservoir at resonance
with an rf field. Pure spin-current injection into low-density
conductors should allow experimental studies of spintronic phenomena
in mesoscopic, ballistic, and nanoscale systems, which up to now has
been largely a playground of theoreticians like Datta and
Das,\cite{Datta:apl90} whose spin transistor concept has stimulated
much of the present interest in spintronics.

The combination of a ferromagnet at the ferromagnetic resonance (FMR)
in Ohmic contact with a
conductor can be interpreted as a
\textquotedblleft spin battery\textquotedblright, with analogies and
differences with charge batteries. For example, charge-current
conservation dictates that a charge battery has two poles, plus and
minus. A spin battery requires only one pole, since the spin current
does not need to be conserved. Furthermore, the polarity is not a
binary, but a three-dimensional vector. The important parameters of a
charge battery are the maximum voltage in the absence of a load, as
well as the maximum charge current, which can be drawn from it. In the
following we present estimates for the analogous characteristics of
the spin battery.

\begin{figure}[ptb]
\includegraphics[scale=0.4]{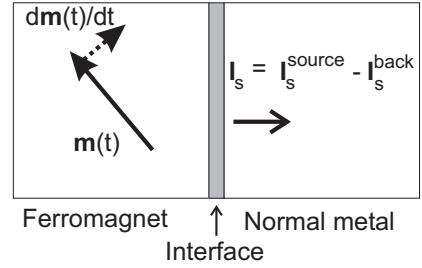}\caption{\label{f1} Schematic
view of the spin battery. Precession of the magnetization
$\mathbf{m}(t)$ of the ferromagnet \textit{F} emits a spin current
$\mathbf{I}_{s}^{\text{source}}$ into the adjacent normal-metal layer
\textit{N}. The spin accumulation in the normal metal either relaxes
by spin-flip scattering or flows back into the ferromagnet, resulting
in a net spin current
$\mathbf{I}_{s}=\mathbf{I}_{s}^{\text{source}}-\mathbf{I}_{s}^{\text{back}}$.}
\end{figure}

Central to our concept is a precessing ferromagnet, which acts as a
source of spin angular momentum, when in contact with normal
metals,\cite{Tserkovnyak:prl02} see Fig.\ \ref{f1}.
This spin injection can be formulated in analogy with the adiabatic
pumping of charge in mesoscopic systems.
\cite{Brataas:prl00,Brouwer:prb98} When the ferromagnet is thicker
than the ferromagnetic coherence length (a few \AA ngstr\o ms in
transition metals such as Co, Ni or Fe), the spin current emitted into
the normal metal is determined by the \textit{mixing conductance}
\cite{Brataas:prl00} $g_{\uparrow \downarrow }=\sum_{nm}[\delta
_{nm}-r_{nm}^{\uparrow } (r_{nm}^{\downarrow })^{\ast}]$ in terms of
the spin-dependent reflection amplitudes $r_{nm}^{\uparrow(\downarrow)
}$ between transverse modes $m$ and $n$ in the normal metal at the
interface to the ferromagnet, where the latter is characterized by the
magnetization direction $\mathbf{m}$. The mixing conductance governs
the transport of spins that are noncollinear to the magnetization
direction in the ferromagnet \cite{Brataas:prl00,Huertas:prb00} and is
also a material parameter proportional to the torque acting on the
ferromagnet in the presence of a noncollinear spin accumulation in the
normal
metal.\cite{Tserkovnyak:prl02,Brataas:prl00,Xia:prb02,Stiles:cm02} For
most systems (with the exception of, \textit{e.g.}, ferromagnetic
insulators \cite{Huertas:prl02}) the imaginary part of the mixing
conductance can be disregarded due to the randomization of phases of
spin-up and spin-down electrons in reciprocal space \cite{Xia:prb02}
and this is assumed in the following. The spin current emitted into
the normal metal is then, simply, \cite{Tserkovnyak:prl02}
\begin{equation}
\mathbf{I}_{s}^{\text{source}}=\frac{\hbar }{4\pi }g_{\uparrow
\downarrow } \mathbf{m}\times \frac{d \mathbf{m}}{dt}\,.
\label{spinemission}
\end{equation}
In our notation, the spin current is measured in units of mechanical
torque. Eq.\ (\ref{spinemission}) is a time-dependent correction to
the Landauer-B\"{u}ttiker formula for noncollinear
ferromagnetic--normal-metal (\textit{F-N}) hybrid
systems.\cite{Brataas:prl00} A simple physical picture can be inferred
from the following thought experiment.\cite{Siegmann:priv} Suppose we
have a \textit{F-N} interface at equilibrium and switch the
magnetization instantaneously. The mismatch of the spin-up and
spin-down chemical potentials leads to large non-equilibrium spin
currents on the length scale of the spin-diffusion length. A slower
magnetization reversal naturally induces smaller spin
currents. Eq. (1) represents the adiabatic limit of the spin currents
pumped by a slow magnetization dynamics. When the spin current
(\ref{spinemission}) is channeled off sufficiently rapidly, the
corresponding loss of angular momentum increases the (Gilbert) damping
of the magnetization dynamics.\cite{Tserkovnyak:prl02} Eq.\
(\ref{spinemission}) is the maximum spin current that can be drawn
from the spin battery.

Next, we need the maximum spin bias obtained when the load
vanishes. When the spin-flip relaxation rate is smaller than the
spin-injection rate, a spin angular momentum $\mathbf{s}$ (in units of
$\hbar$) builds up in the normal metal. We can neglect spatical
dependence within the ferromagnet when the film is sufficiently
thin.\cite{thin} Under these conditions, one finds that the component
of the backflow spin current $\mathbf{I}_s^{\text{back}}$, from the
normal metal to the ferromagnet, parallel to the instantenous
magnetization direction $\mathbf{m}$ is cancelled by an opposite flow
from the ferromagnet. The componenent of $\mathbf{I}_s^{\text{back}}$
perpendicular to $\mathbf{m}$ is\cite{Brataas:prl00}
\begin{equation}
\mathbf{I}_{s}^{\text{back}}=\frac{g_{\uparrow \downarrow}}{2\pi N}\left[\mathbf{s} -  \mathbf{m}\left( \mathbf{m} \cdot
\mathbf{s}\right) \right] \,,
\label{spinback}
\end{equation}
where $N$ is the one-spin density of states.  We note that the mixing
conductance in Eqs.~(\ref{spinemission}) and ~(\ref{spinback}) ought
to be renormalized in highly transparent junctions.\cite{Bauer:future}

The relation between spin excess $\mathbf{s}$ and total spin current
$\mathbf{I}_{s}=\mathbf{I}_s^{\text{source}}-\mathbf{I}_s^{\text{back}}$
in a normal diffusive metal is governed by the spin-diffusion equation
\cite{Johnson:prb88}
\begin{equation}
\frac{\partial \mathbf{s}}{\partial t} = D \frac{\partial^2
\mathbf{s}}{\partial x^2} - \frac{\mathbf{s}}{\tau_{\text{s}}} \, ,
\label{spindiffusion}
\end{equation}
where $D$ is the diffusion coefficient, in three (two) dimensions
$D=v_F^2 \tau/6$ ($D=v_F^2 \tau/4$), and $\tau$, $\tau_{\text{s}}$ are
the elastic and spin-flip relaxation times, respectively. We solve the
diffusion equation with boundary conditions at $x=0$, where $(DA\hbar)
\partial_x \mathbf{s} = -\mathbf{I}_s$ and at the end of the sample
$x=L$, where the spin current vanishes, $\partial_x \mathbf{s}=0$. $A$
is the cross section of the system.

The precession of the magnetization vector of a ferromagnet under a
resonant rf electromagnetic field applied perpendicularly to a DC
magnetic field \cite{Slichter} can be used to drive the spin
battery. The magnitude of the spin current $
\mathbf{I}_{s}^{\text{source}}$ and spin bias $ \Delta
\text{\boldmath$\mu $} \equiv 2 \langle \mathbf{s} \rangle_t /N$ as a
function of the applied field $\mathbf{H}_{0}$ follows from the
Landau-Lifshitz-Gilbert equation $\mathbf{\dot{m}}=-\gamma ^{\ast
}\mathbf{m}\times \mathbf{H}_0 +\alpha \mathbf{m}\times
\mathbf{\dot{m},}$ where $\gamma ^{\ast }$ is the gyromagnetic ratio,
$\alpha \sim 0.01-0.001$ the Gilbert damping factor, and magnetic
anisotropies have been disregarded for simplicity. The spin bias also
has AC components. However, its frequency $\omega$ harmonics are
strongly suppressed when $l_{\text{s}}/(\omega
\tau_{\text{s}})^{1/2}<L<l_{\text{s}}$, which can be easily realized
when $\omega \tau_{\text{s}}>1$, \textit{e.g.}, $\tau_{\text{s}} >
\omega^{-1} \sim 10^{-11} s /
H_0~[\text{T}]$. $l_{\text{s}}=\sqrt{D\tau_{\text{s}}}$ is the
spin-diffusion length in the normal metal. The dominant contribution
to the spin bias is then constant in time and directed along ${\bf
H}_0$. The magnitude of the time-averaged spin accumulation $\Delta
\text{\boldmath$\mu$} \equiv 2 \langle\mathbf{s}(t)\rangle_t/N$ in the
normal metal close to the \textit{F-N} interface then reads:
\begin{equation}
\Delta \mu = \hbar \omega_0 \frac{\sin^2\theta}{\sin^2\theta+\eta}
\, ,
\label{spinaccumulation}
\end{equation}
where the precession cone angle between $\mathbf{H}_{0}$ and
$\mathbf{m}$ is $\theta$, $\eta = (\tau_{\text{i}}/\tau_{\text{s}})
\tanh(L/l_{\text{s}})/(L/l_{\text{s}})$ is a reduction factor, and we
have introduced the spin-injection rate $\tau
_{\text{i}}^{-1}=g_{\uparrow \downarrow }/(2\pi \hbar NAL)$. Large
systems have a smaller injection rate since more states have to be
filled.

The ratio of the injection and spin-flip relaxation times can be
evaluated for a planar geometry. We consider a free-electron gas in
contact with a metallic ferromagnet. The mixing conductance is
$g_{\uparrow \downarrow}=\kappa A k_F^2/(4\pi)$ ($g_{\uparrow
\downarrow}=\kappa A k_F/\pi$) for spin injection into
three-(two-)dimensional systems. First-principles band-structure
calculations show that for combinations like Co/Cu or Fe/Cr $\kappa$
remains close to unity.\cite{Xia:prb02} The ratio between the
injection and spin-flip relaxation times in three [two] dimensions can
be calculated to be $\tau_{\text{i}}/\tau_{\text{s}}=
\sqrt{8/3}\kappa^{-1} \sqrt{\epsilon} (L/l_{\text{s}})$
[$\tau_{\text{i}}/\tau_{\text{s}}=2\kappa^{-1} \sqrt{\epsilon}
(L/l_{\text{s}})$]. $\epsilon=\tau/\tau_{\text{s}}$ is the ratio of
the elastic scattering rate and the spin-flip relaxation rate, which
is usually much smaller than unity.

When the spin relaxation time is longer than the spin injection time
and the precession cone angle is sufficiently large, $\sin^2{\theta} >
\eta$, the spin bias saturates at its maximum value $\Delta
\mu_0=\hbar \omega_0$. In this regime the spin accumulation does not
depend on the material parameters. It should be feasible to realize
the full spin bias when $L \ll l_s$ since $\eta \approx \sqrt{8/3}
\kappa^{-1} \sqrt{\epsilon} (L/l_s)$, \textit{e.g.} when $L/l_s=0.1$,
$\sqrt{8/3} \kappa^{-1} \sqrt{\epsilon}=0.1$ the precession cone angle
should be larger than 6 degrees. For small precession cone angles
$\theta \approx H_1/(\alpha H_0)$, so for \textit{e.g.}  $H_0=1.0$ T,
$\alpha=10^{-3}$ this requires a $H_1=0.1 $ mT rf field with a
resulting spin bias of $\Delta \mu=0.1 meV$. For a smaller precession
angle, \textit{e.g.}  $\theta=0.6$ degrees the spin-bias is smaller,
$\Delta \mu = 1 \mu V$, but still clearly measurable. Epitaxially
grown clean samples with even longer spin-diffusion lengths and
smaller spin-flip to non-spin flip relaxation ratios $\epsilon$ will
function as spin-batteries with smaller precession angles. The
precession cone angle $\theta$ in FMR is typically small, but
$\theta>$ 15 degrees can be achieved for a sufficiently intense rf
field and a soft ferromagnet, \textit{e.g.},
permalloy.\cite{Heinrich:priv} The maximum DC spin current source is
$|\langle\mathbf{I}_{s}^{\text{source}}\rangle_t|\approx \hbar
\omega_0 \kappa Ak_F^2 \sin^2\theta/(4 \pi)$, \textit{e.g.} for a
precession cone angle of $\theta=$ 6 degrees the equivalent electrical
spin current $(e/\hbar)|\langle \mathbf{I}_{s}^{\text{source}}
\rangle_t|$ is 0.1 nano Ampere per conducting channel $A
k_F^2/(4\pi)$.  The total number of channels is a large number since
the cross sections may be chosen very much larger than the Fermi
wavelength thus ensuring that a large spin current may be drawn from
the battery.

Ferromagnetic resonance dissipates energy proportional to the damping
parameter $\alpha$ of the magnetization dynamics. The power loss
$dE/dt= \alpha \hbar \omega_0^2 N_s \sin^2\theta$ is proportional to
the volume of the ferromagnet through the number of spins in the
ferromagnet in units of $\hbar$, $N_s$. The power loss can be
significant even for a thin film ferromagnet, \textit{e.g.}, for a 10
monolayer thick Fe film with $\alpha \sim 10^{-3}$, $\sin^2{\theta}
\sim 10^{-2}$, and $\omega_0 \sim 10^{11}$~s$^{-1}$, the power loss
per unit area is $(1/A) dE/dt \sim 0.1$~W/cm$^2$. The temperature can
be kept low by \textit{e.g.} immersing the sample in superfluid helium.
The heat transfer is then approximately $8$~W/cm$^2$K for small
temperature gradients and increases for larger temperature
gradients,\cite{Skocpol:jap74} which appears sufficient for the
present purpose.

Schmidt \textit{et al.}  \cite{Schmidt:prb00} realized that efficient
spin injection into semiconductors by Ohmic contacts is close to
impossible with transition-metal ferromagnets since virtually all of
the applied potential drops over the nonmagnetic part and is wasted
for spin injection. The present mechanism does not rely on an applied
bias and does not suffer from the conductance mismatch, because the
smallness of the mixing conductance for a ferromagnet-semiconductor
interface is compensated by the small spin current that is necessary
to saturate the spin accumulation.

Possible undesirable spin precession and energy generation in the
normal-metal parts of the system is of no concern for material
combinations with different $g$-factors, as, \textit{e.g.}, Fe
($g=2.1$) and GaAs ($g=-0.4$), or when the magnetic anisotropy 
modifies the resonance frequency with respect to electrons
in the normal metal.

The optimal material combinations for a battery depend on the planned
usage. From Eq.\ (\ref{spinemission}) it follows that the largest spin
current can be achieved when the conductor is a normal metal, whereas any 
material combination appears suitable when the load is small, as long as the 
contact is Ohmic and the system
is smaller than the spin-diffusion length. 

Standard metals, like Al and Cu, are good candidate materials, since
the spin-diffusion length is very long, $l_{\text{s}}\sim1~\mu$m at
low temperatures, and remains quite long at room
temperature.\cite{Jedema:nat01, Jedema:nat02} Indirect indications of
spin accumulation in Cu can be deduced from the absence of any
enhancement of the Gilbert damping in FMR when in contact with thin
ferromagnetic films.\cite{Mizukami:mmm01, Tserkovnyak:prl02}

Semiconductors have the advantage of a larger ratio of spin bias to
Fermi energy. Let us first consider the case of GaAs. The spin-flip
relaxation time in GaAs can be very long, 
$\tau_{\text{s}}=10^{-7}~\text{s}$ at
$n=5 \times 10^{16}$~cm$^{3}$
carrier density.\cite{Kawakami:sci01} These favorable numbers are offset by the 
difficulty to form
Ohmic contacts to GaAs, however. Large Schottky barrier exponentially
suppress the interfaces mixing conductance parameter $\kappa$.

InAs has the advantage of a natural accumulation layer at the surface
that avoids Schottky barriers when covered by high-density
metals. However, the spin-orbit interaction in a narrow gap
semiconductor like InAs is substantial, which reduces
$\tau_{\text{s}}$. In asymmetric confinement structures, the spin-flip
relaxation rate is governed by the Rashba type spin-orbit interaction,
which vanishes in symmetrical quantum wells.\cite{Nitta:prl97} The
remaining D'yakonov-Perel scattering rate is reduced in narrow
quasi-one-dimensional channels of width $d$ due to waveguide diffusion
modes by a factor of $(L_s/d)^2$, where $L_s=v_F/|h(k_F)|$ is the
spin-precession length in terms of the spin-orbit coupling
$H_{so}=\mathbf{h}(\mathbf{k}) \cdot
\mathbf{s}$,\cite{Malshukov:prb00} which makes InAs based systems an
interesting material for a spin battery as well.

In Si, the spin-flip relaxation time is long, since spin-orbit
interaction is weak. Furthermore, the possibility of heavy doping
allows control of Schottky barriers. So, Si appears to be a good
candidate for spin injection into semiconductors.

The spin bias can be detected noninvasively \textit{via} tunnel
junctions with an analyzing ferromagnet having a switchable
magnetization direction. A voltage difference of $p\Delta\mu$ is
detected for parallel and anti-parallel configurations of the
analyzing magnetization with respect to the spin accumulation in the
normal metal, where
$p=(G_{\uparrow}-G_{\downarrow})/(G_{\uparrow}+G_{\downarrow})$ is the
relative polarization of the tunnel conductances of the contact. The
test magnetic layer need not be flipped. It is sufficient to reverse
the direction of the DC static magnetic field. The spin current, on the
other hand, can be measured \textit{via} the drop of spin bias over a
known resistive element.

Spin-pumping into the normal metal can also have consequences for the
nuclei \textit{via} the hyperfine interaction between electrons and
nuclear spins.\cite{Kawakami:sci01} An
initially unpolarized collection of nuclear spins can be oriented by a
spin-polarized electron current, which transfers angular momentum by
spin-flop scattering. A ferromagnetically ordered nuclear-spin system
can lead to an Overhauser\cite{ Overhauser:pr53} field on the electron
spin. This effect does not affect the spin bias $\Delta
\text{\boldmath$\mu$}$, but induces an equilibrium spin density in the
normal metal $\mathbf{s}^0$ \textit{via} the nuclear magnetic field,
and can be exploited in experiments where the the total spin-density
$\mathbf{s}+\mathbf{s}_0$ is an important parameter. The
electron-nuclear interaction can be included by adding
\cite{Overhauser:pr53,Slichter}
\begin{equation}
{\bf I}_s^{\text{nuc}}=\frac{\hbar{\bf s}_n}{T_n}
\label{Inuc}
\end{equation}
to the electron spin dynamics so that ${\bf I} \rightarrow {\bf
I}_s^{\text{source}}-{\bf I}_s^{\text{back}}+{\bf I}_s^{\text{nuc}}$,
where ${\bf s}_{n}$ is the nonequilibrium nuclear spin accumulation
and $T_n$ is the electron-nuclear relaxation time. The nuclear spin
dynamics is described by
\begin{equation}
\frac{d {\bf s}_n}{dt} = - \frac{{\bf s}_n}{T_n'}+ \frac{{\bf s}}{T_e} \, ,
\end{equation}
where $T_n' \leq T_n$ is the nuclear-spin relaxation time and $T_e$ is
the nuclear-electron relaxation. In steady state,
$\mathbf{s}_n=(T_n'/T_n)(T_n/T_e)\mathbf{s}$. In the experimentally
relevant regime $T_e^{-1} \ll \tau_i^{-1}$ the electron-nuclear
interaction (\ref{Inuc}) has a negligible effect on the nonequilibrium
spin accumulation $\mathbf{s}$ and thus Eq.~(\ref{spinaccumulation})
remains unchanged. $T_n/T_e=8I(I+1) \epsilon_F n_N /(9 k_B T n_e)$ for
small polarizations, where $\epsilon_F$ is the Fermi energy of the
electron gas, $k_BT$ is the thermal energy, $n_N$ is the nuclear
density and $n_e$ is the one-spin electron
density.\cite{Overhauser:pr53} Using $N=(3/2) n_e/\epsilon_F$
($N=n_e/\epsilon_F$ in two dimensions) and
Eq.~(\ref{spinaccumulation}) the relative enhancement of the DC
nuclear spin polarization is
\begin{equation}
{\bf s}_n = n_N \frac{T_n'}{T_n} \frac{2}{3} I (I+1)
\frac{\Delta \text{\boldmath$ \mu$}}{k_B T}   \, .
\end{equation}
for $\Delta \mu \ll k_BT$.  The nuclear-spin
polarization increases with the spin bias and by lowering the
temperature. The hyperpolarized nuclei, in turn, produce an effective
nuclear field that polarizes the {\em equilibrium} properties of the
electron gas $\mathbf{s}^0$. In bulk GaAs, the nuclear magnetic field
is $B_n=5.3$~T when the nuclei are fully spin-polarized which should
occur at sufficiently low temperatures.\cite{Paget:prb77}

Berger\cite{Berger:prb99} proposed to generate a DC voltage by the
FMR, which bears similarities with our proposal. But Berger's
mechanism of spin injection, originating from the spin-flip scattering
in the ferromagnet as induced by spin waves appears to be different
from ours. We propose to achieve spin injection \textit{via} the
modulation of the interface scattering matrix by the coherent
precession of the magnetization, which allows, for example,
quantitative calculations for various materials.

In conclusion, we present the new concept of a spin battery, which is
a source of spin, just as a conventional battery is a source of
charge, and estimate its performance for different
material combinations.

We are grateful to I.\ Appeli, B.\ Heinrich, A.~D.\ Kent, D.\ Monsma,
and H.~C.\ Siegmann for stimulating discussions. This work was
supported in part by the DARPA award No. MDA 972-01-1-0024, the NEDO
International Joint Research Grant Program \textquotedblleft
Nano-magnetoelectronics\textquotedblright, FOM, the Schlumberger
Foundation, and NSF Grant No. DMR 99-81283. G.\ E.\ W.\ B.\
acknowledges the hospitality of Dr.\ Y.\ Hirayama and his group at the
NTT Basic Research Laboratories.


\begin{thebibliography}{36}
\expandafter\ifx\csname natexlab\endcsname\relax\def\natexlab#1{#1}\fi
\expandafter\ifx\csname bibnamefont\endcsname\relax
\def\bibnamefont#1{#1}\fi \expandafter\ifx\csname
bibfnamefont\endcsname\relax \def\bibfnamefont#1{#1}\fi
\expandafter\ifx\csname citenamefont\endcsname\relax
\def\citenamefont#1{#1}\fi \expandafter\ifx\csname url\endcsname\relax
\def\url#1{\texttt{#1}}\fi \expandafter\ifx\csname
urlprefix\endcsname\relax\def\urlprefix{URL }\fi
\providecommand{\bibinfo}[2]{#2}
\providecommand{\eprint}[2][]{\url{#2}}

\bibitem[{\citenamefont{Wolf et~al.}(2001)\citenamefont{Wolf,
Awschalom, Buhrman, Daughton, von Molnar, Roukes, Chtchelkanova, and
Treger}}]{wolf:sci01}
\bibinfo{author}{\bibfnamefont{S.}~\bibnamefont{Wolf}},
\bibinfo{author}{\bibfnamefont{D.~D.} \bibnamefont{Awschalom}},
\bibinfo{author}{\bibfnamefont{R.}~\bibnamefont{Buhrman}},
\bibinfo{author}{\bibfnamefont{J.}~\bibnamefont{Daughton}},
\bibinfo{author}{\bibfnamefont{S.}~\bibnamefont{von Molnar}},
\bibinfo{author}{\bibfnamefont{M.}~\bibnamefont{Roukes}},
\bibinfo{author}{\bibfnamefont{A.}~\bibnamefont{Chtchelkanova}},
\bibnamefont{and}
\bibinfo{author}{\bibfnamefont{D.}~\bibnamefont{Treger}},
\bibinfo{journal}{Science} \textbf{\bibinfo{volume}{294}},
\bibinfo{pages}{1488} (\bibinfo{year}{2001}).

\bibitem[{\citenamefont{Baibich et~al.}(1988)\citenamefont{Baibich,
Broto, Fert, Dau, Petroff, Eitenne, Creuzet, Friederich, and
Chazelas}}]{Baibich:prl88}
\bibinfo{author}{\bibfnamefont{M.}~\bibnamefont{Baibich}},
\bibinfo{author}{\bibfnamefont{J.~M.} \bibnamefont{Broto}},
\bibinfo{author}{\bibfnamefont{A.}~\bibnamefont{Fert}},
\bibinfo{author}{\bibfnamefont{F.~N.~V.} \bibnamefont{Dau}},
\bibinfo{author}{\bibfnamefont{F.}~\bibnamefont{Petroff}},
\bibinfo{author}{\bibfnamefont{P.}~\bibnamefont{Eitenne}},
\bibinfo{author}{\bibfnamefont{G.}~\bibnamefont{Creuzet}},
\bibinfo{author}{\bibfnamefont{A.}~\bibnamefont{Friederich}},
\bibnamefont{and}
\bibinfo{author}{\bibfnamefont{J.}~\bibnamefont{Chazelas}},
\bibinfo{journal}{Phys. Rev. Lett.} \textbf{\bibinfo{volume}{61}},
\bibinfo{pages}{2472} (\bibinfo{year}{1988}).

\bibitem[{\citenamefont{Miyazaki and Tezuka}(1995)}]{Miyazaki:jmmm95}
\bibinfo{author}{\bibfnamefont{T.}~\bibnamefont{Miyazaki}}
\bibnamefont{and}
\bibinfo{author}{\bibfnamefont{N.}~\bibnamefont{Tezuka}},
\bibinfo{journal}{J. Magn. Magn. Mater.}
\textbf{\bibinfo{volume}{139}}, \bibinfo{pages}{L231}
(\bibinfo{year}{1995}); \bibinfo{author}{\bibfnamefont{J.~S.}
\bibnamefont{Moodera}}, \bibinfo{author}{\bibfnamefont{L.~R.}
\bibnamefont{Kinder}}, \bibinfo{author}{\bibfnamefont{T.~M.}
\bibnamefont{Wong}}, \bibnamefont{and}
\bibinfo{author}{\bibfnamefont{R.}~\bibnamefont{Meservey}},
\bibinfo{journal}{Phys. Rev. Lett.} \textbf{\bibinfo{volume}{74}},
\bibinfo{pages}{3273} (\bibinfo{year}{1995}).`

\bibitem[{\citenamefont{Schmidt et~al.}(2000)\citenamefont{Schmidt,
Ferrand, Molenkamp, Filip, and van Wees}}]{Schmidt:prb00}
\bibinfo{author}{\bibfnamefont{G.}~\bibnamefont{Schmidt}},
\bibinfo{author}{\bibfnamefont{D.}~\bibnamefont{Ferrand}},
\bibinfo{author}{\bibfnamefont{L.~W.} \bibnamefont{Molenkamp}},
\bibinfo{author}{\bibfnamefont{A.~T.} \bibnamefont{Filip}},
\bibnamefont{and} \bibinfo{author}{\bibfnamefont{B.~J.}
\bibnamefont{van Wees}}, \bibinfo{journal}{Phys. Rev. B}
\textbf{\bibinfo{volume}{62}}, \bibinfo{pages}{R4790}
(\bibinfo{year}{2000}).

\bibitem[{\citenamefont{Kikkawa and Awschalom}(1999)}]{Kikkawa:nat99}
\bibinfo{author}{\bibfnamefont{J.~M.} \bibnamefont{Kikkawa}}
\bibnamefont{and} \bibinfo{author}{\bibfnamefont{D.~D.}
\bibnamefont{Awschalom}}, \bibinfo{journal}{Nature}
\textbf{\bibinfo{volume}{397}}, \bibinfo{pages}{139}
(\bibinfo{year}{1999}).

\bibitem[{\citenamefont{Fiederling
et~al.}(1999)\citenamefont{Fiederling, Keim, Reuscher, Ossau, Schmidt,
Waag, and Molenkamp}}]{Fieder:nat99}
\bibinfo{author}{\bibfnamefont{R.}~\bibnamefont{Fiederling}},
\bibinfo{author}{\bibfnamefont{M.}~\bibnamefont{Keim}},
\bibinfo{author}{\bibfnamefont{G.}~\bibnamefont{Reuscher}},
\bibinfo{author}{\bibfnamefont{W.}~\bibnamefont{Ossau}},
\bibinfo{author}{\bibfnamefont{G.}~\bibnamefont{Schmidt}},
\bibinfo{author}{\bibfnamefont{A.}~\bibnamefont{Waag}},
\bibnamefont{and} \bibinfo{author}{\bibfnamefont{L.~W.}
\bibnamefont{Molenkamp}}, \bibinfo{journal}{Nature}
\textbf{\bibinfo{volume}{402}}, \bibinfo{pages}{787}
(\bibinfo{year}{1999});
\bibinfo{author}{\bibfnamefont{Y.}~\bibnamefont{Ohno}},
\bibinfo{author}{\bibfnamefont{D.~K.} \bibnamefont{Young}},
\bibinfo{author}{\bibfnamefont{B.}~\bibnamefont{Beschoten}},
\bibinfo{author}{\bibfnamefont{F.}~\bibnamefont{Matsukura}},
\bibinfo{author}{\bibfnamefont{H.}~\bibnamefont{Ohno}},
\bibnamefont{and} \bibinfo{author}{\bibfnamefont{D.~D.}
\bibnamefont{Awschalom}}, \bibinfo{journal}{Nature}
\textbf{\bibinfo{volume}{402}}, \bibinfo{pages}{790}
(\bibinfo{year}{1999}).

\bibitem[{\citenamefont{Monsma et~al.}(1998)\citenamefont{Monsma,
Vlutters, and Lodder}}]{Monsma:Sc98}
\bibinfo{author}{\bibfnamefont{D.}~\bibnamefont{Monsma}},
\bibinfo{author}{\bibfnamefont{R.}~\bibnamefont{Vlutters}},
\bibnamefont{and}
\bibinfo{author}{\bibfnamefont{J.}~\bibnamefont{Lodder}},
\bibinfo{journal}{Science} \textbf{\bibinfo{volume}{281}},
\bibinfo{pages}{407} (\bibinfo{year}{1998});
\bibinfo{author}{\bibfnamefont{H.~J.} \bibnamefont{Zhu}},
\bibinfo{author}{\bibfnamefont{M.}~\bibnamefont{Ramsteiner}},
\bibinfo{author}{\bibfnamefont{H.}~\bibnamefont{Kostial}},
\bibinfo{author}{\bibfnamefont{M.}~\bibnamefont{Wassermeier}},
\bibinfo{author}{\bibfnamefont{H.-P.} \bibnamefont{Schonherr}},
\bibnamefont{and} \bibinfo{author}{\bibfnamefont{K.~H.}
\bibnamefont{Ploog}}, \bibinfo{journal}{Phys. Rev. Lett.}
\textbf{\bibinfo{volume}{87}}, \bibinfo{pages}{016601}
(\bibinfo{year}{2001}); \bibinfo{author}{\bibfnamefont{A.~T.}
\bibnamefont{Hanbicki}}, \bibinfo{author}{\bibfnamefont{B.~T.}
\bibnamefont{Jonker}},
\bibinfo{author}{\bibfnamefont{G.}~\bibnamefont{Itskos}},
\bibinfo{author}{\bibfnamefont{G.}~\bibnamefont{Kioseoglou}},
\bibnamefont{and}
\bibinfo{author}{\bibfnamefont{A.}~\bibnamefont{Petrou}}
(\bibinfo{year}{2001}), \bibinfo{note}{cond-mat/0110059};
\bibinfo{author}{\bibfnamefont{V.}~\bibnamefont{Motsnyi}},
\bibinfo{author}{\bibfnamefont{V.}~\bibnamefont{Safarov}},
\bibinfo{author}{\bibfnamefont{J.~D.} \bibnamefont{Boeck}},
\bibinfo{author}{\bibfnamefont{W.~v.~R.} \bibnamefont{J.~Das}},
\bibinfo{author}{\bibfnamefont{E.}~\bibnamefont{Goovaerts}},
\bibnamefont{and}
\bibinfo{author}{\bibfnamefont{G.}~\bibnamefont{Borghs}}
(\bibinfo{year}{2001}), \bibinfo{note}{cond-mat/0110240};
\bibinfo{author}{\bibfnamefont{S.}~\bibnamefont{Parkin}},
\bibinfo{journal}{talk at ICMFS} (\bibinfo{year}{2002}).

\bibitem[{\citenamefont{Jedema et~al.}(2001)\citenamefont{Jedema,
Filip, and van Wees}}]{Jedema:nat01}
\bibinfo{author}{\bibfnamefont{F.~F.} \bibnamefont{Jedema}},
\bibinfo{author}{\bibfnamefont{A.~T.} \bibnamefont{Filip}},
\bibnamefont{and} \bibinfo{author}{\bibfnamefont{B.~J.}
\bibnamefont{van Wees}}, \bibinfo{journal}{Nature}
\textbf{\bibinfo{volume}{410}}, \bibinfo{pages}{345}
(\bibinfo{year}{2001}).

\bibitem[{\citenamefont{Datta and Das}(1990)}]{Datta:apl90}
\bibinfo{author}{\bibfnamefont{S.}~\bibnamefont{Datta}}
\bibnamefont{and}
\bibinfo{author}{\bibfnamefont{B.}~\bibnamefont{Das}},
\bibinfo{journal}{Appl. Phys. Lett.} \textbf{\bibinfo{volume}{56}},
\bibinfo{pages}{665} (\bibinfo{year}{1990}).

\bibitem[{\citenamefont{Tserkovnyak
et~al.}(2002)\citenamefont{Tserkovnyak, Brataas, andBauer}}]{Tserkovnyak:prl02}
\bibinfo{author}{\bibfnamefont{Y.}~\bibnamefont{Tserkovnyak}},
\bibinfo{author}{\bibfnamefont{A.}~\bibnamefont{Brataas}},
\bibnamefont{and} \bibinfo{author}{\bibfnamefont{G.~E.~W.}
\bibnamefont{Bauer}}, \bibinfo{journal}{Phys. Rev. Lett.}
\textbf{\bibinfo{volume}{88}}, \bibinfo{pages}{117601}
(\bibinfo{year}{2002}).

\bibitem[{\citenamefont{Brouwer}(1998)}]{Brouwer:prb98}
\bibinfo{author}{\bibfnamefont{P.~W.} \bibnamefont{Brouwer}},
\bibinfo{journal}{Phys. Rev. B} \textbf{\bibinfo{volume}{58}},
\bibinfo{pages}{R10135} (\bibinfo{year}{1998}).

\bibitem[{\citenamefont{Brataas et~al.}(2000)\citenamefont{Brataas,
Nazarov, and Bauer}}]{Brataas:prl00}
\bibinfo{author}{\bibfnamefont{A.}~\bibnamefont{Brataas}},
\bibinfo{author}{\bibfnamefont{Y.~V.} \bibnamefont{Nazarov}},
\bibnamefont{and} \bibinfo{author}{\bibfnamefont{G.~E.~W.}
\bibnamefont{Bauer}}, \bibinfo{journal}{Phys. Rev. Lett.}
\textbf{\bibinfo{volume}{84}}, \bibinfo{pages}{2481}
(\bibinfo{year}{2000}); \bibinfo{journal}{Eur. Phys. J. B}
\textbf{\bibinfo{volume}{22}}, \bibinfo{pages}{99}
(\bibinfo{year}{2001}).

\bibitem[{\citenamefont{Huertas-Hernando
et~al.}(2000)\citenamefont{Huertas-Hernando, Nazarov, Brataas, and
Bauer}}]{Huertas:prb00}
\bibinfo{author}{\bibfnamefont{D.}~\bibnamefont{Huertas-Hernando}},
\bibinfo{author}{\bibfnamefont{Y.~V.} \bibnamefont{Nazarov}},
\bibinfo{author}{\bibfnamefont{A.}~\bibnamefont{Brataas}},
\bibnamefont{and} \bibinfo{author}{\bibfnamefont{G.~E.~W.}
\bibnamefont{Bauer}}, \bibinfo{journal}{Phys. Rev. B}
\textbf{\bibinfo{volume}{62}}, \bibinfo{pages}{5700}
(\bibinfo{year}{2000}).

\bibitem[{\citenamefont{Xia
et~al.}(2001{\natexlab{a}})\citenamefont{Xia, Kelly, Bauer, Brataas,
and Turek}}]{Xia:prb02}
\bibinfo{author}{\bibfnamefont{K.}~\bibnamefont{Xia}},
\bibinfo{author}{\bibfnamefont{P.~J.} \bibnamefont{Kelly}},
\bibinfo{author}{\bibfnamefont{G.~E.~W.} \bibnamefont{Bauer}},
\bibinfo{author}{\bibfnamefont{A.}~\bibnamefont{Brataas}},
\bibnamefont{and}
\bibinfo{author}{\bibfnamefont{I.}~\bibnamefont{Turek}},
\bibinfo{journal}{Phys. Rev. B} \textbf{\bibinfo{volume}{65}},
\bibinfo{pages}{220401(R)} (\bibinfo{year}{2002}).


\bibitem[{\citenamefont{Stiles and Zangwill}(2002)}]{Stiles:cm02}
\bibinfo{author}{\bibfnamefont{M.~D.} \bibnamefont{Stiles}}
\bibnamefont{and}
\bibinfo{author}{\bibfnamefont{A.}~\bibnamefont{Zangwill}}
(\bibinfo{year}{2002}), \bibinfo{note}{cond-mat/0202397}.

\bibitem[{\citenamefont{Huertas-Hernando
et~al.}(2002)\citenamefont{Huertas-Hernando, Nazarov, and
Belzig}}]{Huertas:prl02}
\bibinfo{author}{\bibfnamefont{D.}~\bibnamefont{Huertas-Hernando}},
\bibinfo{author}{\bibfnamefont{Y.~V.} \bibnamefont{Nazarov}},
\bibnamefont{and}
\bibinfo{author}{\bibfnamefont{W.}~\bibnamefont{Belzig}},
\bibinfo{journal}{Phys. Rev. Lett.} \textbf{\bibinfo{volume}{88}},
\bibinfo{pages}{047003} (\bibinfo{year}{2002}).

\bibitem[{\citenamefont{Siegmann}()}]{Siegmann:priv}
\bibinfo{title}{This argument is due to H.~C. Siegmann.}

\bibitem[{\citenamefont{nobody}()}]{thin}
\bibinfo{title}{We
require that the ferromagnetic film is thinner than the spin diffusion
length and $\lambda_F (J/\hbar \omega_0)$, where $J$ is the exchange
(stiffness) energy, $\omega_0$ is the rf frequency and $\lambda_F$ is
the Fermi wavelength.}


\bibitem[{\citenamefont{Gerrit E. W.~Bauer and
Brataas}(2002)}]{Bauer:future}
\bibinfo{author}{\bibnamefont{G.~E.~W.}~\bibnamefont{Bauer}},
\bibinfo{author}{\bibfnamefont{Y.}~\bibnamefont{Tserkovnyak}},
\bibinfo{author}{\bibnamefont{D.}~\bibnamefont{Huertas-Hernando}},
\bibinfo{author}{\bibfnamefont{A.}~\bibnamefont{Brataas}}
(\bibinfo{year}{2002}), \bibinfo{note}{cond-mat/0205453};
\bibinfo{author}{\bibfnamefont{K.~M.} \bibnamefont{Schep}},
\bibinfo{author}{\bibfnamefont{J.~B. A.~N.} \bibnamefont{van Hoof}},
\bibinfo{author}{\bibfnamefont{P.~J.} \bibnamefont{Kelly}},
\bibinfo{author}{\bibfnamefont{G.~E.~W.} \bibnamefont{Bauer}},
\bibnamefont{and} \bibinfo{author}{\bibfnamefont{J.~E.}
\bibnamefont{Inglesfield}}, \bibinfo{journal}{Phys. Rev. B}
\textbf{\bibinfo{volume}{56}}, \bibinfo{pages}{10805}
(\bibinfo{year}{1997}).



\bibitem[{\citenamefont{Johnson et~al.}(1988)\citenamefont{Johnson and
Silsbee}}]{Johnson:prb88}
\bibinfo{author}{\bibfnamefont{M.}~\bibnamefont{Johnson}},
\bibinfo{author}{\bibfnamefont{R.~H.}~\bibnamefont{Silsbee}},
\bibinfo{journal}{Phys. Rev. B} \textbf{\bibinfo{volume}{37}},
\bibinfo{pages}{5312} (\bibinfo{year}{2000}).

\bibitem[{\citenamefont{Slichter}(1990)}]{Slichter}
\bibinfo{author}{\bibfnamefont{C.~P.} \bibnamefont{Slichter}},
\emph{\bibinfo{title}{Principles of Magnetic Resonance}}
(\bibinfo{publisher}{Springer-Verlag}, \bibinfo{address}{NY, U.S.},
\bibinfo{year}{1990}), \bibinfo{edition}{3rd} ed.  

\bibitem[{\citenamefont{Heinrich}()}]{Heinrich:priv}
\bibinfo{author}{\bibfnamefont{B.}~\bibnamefont{Heinrich}},
\emph{\bibinfo{title}{private communication}}.

\bibitem[{\citenamefont{Skocpol}(1974)}]{Skocpol:jap74}
\bibinfo{author}{\bibfnamefont{W.~J.} \bibnamefont{Skocpol}},
\bibinfo{author}{\bibfnamefont{M.~R.} \bibnamefont{Beasley}},
\bibinfo{author}{\bibfnamefont{M.}    \bibnamefont{Tinkham}},
\bibinfo{journal}{J. App. Phys.} \textbf{\bibinfo{volume}{45}},
\bibinfo{pages}{4054} (\bibinfo{year}{1974})


\bibitem[{\citenamefont{Jedema et~al.}(2002)\citenamefont{Jedema,
Heershce, Filip, Baselmans, and van Wees}}]{Jedema:nat02}
\bibinfo{author}{\bibfnamefont{F.~J.} \bibnamefont{Jedema}},
\bibinfo{author}{\bibfnamefont{H.~B.} \bibnamefont{Heershce}},
\bibinfo{author}{\bibfnamefont{A.~T.} \bibnamefont{Filip}},
\bibinfo{author}{\bibfnamefont{J.~J.~A.} \bibnamefont{Baselmans}},
\bibnamefont{and} \bibinfo{author}{\bibfnamefont{B.~J.}
\bibnamefont{van Wees}}, \bibinfo{journal}{Nature}
(\bibinfo{year}{2002}), \bibinfo{note}{in press}.

\bibitem[{\citenamefont{Mizukami
et~al.}(2001{\natexlab{a}})\citenamefont{Mizukami, Ando, and
Miyazaki}}]{Mizukami:mmm01}
\bibinfo{author}{\bibfnamefont{S.}~\bibnamefont{Mizukami}},
\bibinfo{author}{\bibfnamefont{Y.}~\bibnamefont{Ando}},
\bibnamefont{and}
\bibinfo{author}{\bibfnamefont{T.}~\bibnamefont{Miyazaki}},
\bibinfo{journal}{J. Magn. Magn. Mater.}
\textbf{\bibinfo{volume}{226}}, \bibinfo{pages}{1640}
(\bibinfo{year}{2001}{\natexlab{a}});
\bibinfo{author}{\bibfnamefont{S.}~\bibnamefont{Mizukami}},
\bibinfo{author}{\bibfnamefont{Y.}~\bibnamefont{Ando}},
\bibnamefont{and}
\bibinfo{author}{\bibfnamefont{T.}~\bibnamefont{Miyazaki}},
\bibinfo{journal}{Jpn. J. Appl. Phys.} \textbf{\bibinfo{volume}{40}},
\bibinfo{pages}{580} (\bibinfo{year}{2001}{\natexlab{b}}).

\bibitem[{\citenamefont{Kawakami et~al.}(2001)\citenamefont{Kawakami,
Kato, Hanson, Malajovich, Stephens, Johnston-Halperin, Salis, Gossard,
and Awschalom}}]{Kawakami:sci01} \bibinfo{author}{\bibfnamefont{R.~K.}
\bibnamefont{Kawakami}},
\bibinfo{author}{\bibfnamefont{Y.}~\bibnamefont{Kato}},
\bibinfo{author}{\bibfnamefont{M.}~\bibnamefont{Hanson}},
\bibinfo{author}{\bibfnamefont{I.}~\bibnamefont{Malajovich}},
\bibinfo{author}{\bibfnamefont{J.~M.} \bibnamefont{Stephens}},
\bibinfo{author}{\bibfnamefont{E.}~\bibnamefont{Johnston-Halperin}},
\bibinfo{author}{\bibfnamefont{G.}~\bibnamefont{Salis}},
\bibinfo{author}{\bibfnamefont{A.~C.} \bibnamefont{Gossard}},
\bibnamefont{and} \bibinfo{author}{\bibfnamefont{D.~D.}
\bibnamefont{Awschalom}}, \bibinfo{journal}{Science}
\textbf{\bibinfo{volume}{294}}, \bibinfo{pages}{131}
(\bibinfo{year}{2001}); \bibinfo{author}{\bibfnamefont}{J.~H.}~\bibnamefont{Smet},
\bibinfo{author}{\bibfnamefont}{R.~A.}~\bibnamefont{Deutschmann},
\bibinfo{author}{\bibfnamefont}{F.}~\bibnamefont{Ertl},
\bibinfo{author}{\bibfnamefont}{W.}~\bibnamefont{Wegscheider},
\bibinfo{author}{\bibfnamefont}{G.}~\bibnamefont{Abstreiter},
\bibinfo{author}{\bibfnamefont}{K.}~\bibnamefont{von Klitzing}
\bibinfo{journal}{Nature} \textbf{\bibinfo{volume}{415}},
\bibinfo{pages}{281} {\bibinfo{year}{2002}}; \bibinfo{author}{\bibfnamefont}{K.}~\bibnamefont{Hashimoto},
\bibinfo{author}{\bibfnamefont}{K.}~\bibnamefont{Muraki},
\bibinfo{author}{\bibfnamefont}{T.}~\bibnamefont{Saku},
\bibinfo{author}{\bibfnamefont}{Y.}~\bibnamefont{Hirayama} \bibinfo{journal}{Phys. 
Rev. Lett.}, \bibinfo{note}{in press}


\bibitem[{\citenamefont{Nitta et~al.}(1997)\citenamefont{Nitta,
Akazaki, Takayanagi, and Enoki}}]{Nitta:prl97}
\bibinfo{author}{\bibfnamefont{J.}~\bibnamefont{Nitta}},
\bibinfo{author}{\bibfnamefont{T.}~\bibnamefont{Akazaki}},
\bibinfo{author}{\bibfnamefont{H.}~\bibnamefont{Takayanagi}},
\bibnamefont{and}
\bibinfo{author}{\bibfnamefont{T.}~\bibnamefont{Enoki}},
\bibinfo{journal}{Phys. Rev. Lett.} \textbf{\bibinfo{volume}{78}},
\bibinfo{pages}{1335} (\bibinfo{year}{1997});
\bibinfo{author}{\bibfnamefont{G.}~\bibnamefont{Engels}},
\bibinfo{author}{\bibfnamefont{J.}~\bibnamefont{Lange}},
\bibinfo{author}{\bibfnamefont{T.}~\bibnamefont{Sch{\"{a}}pers}},
\bibnamefont{and}
\bibinfo{author}{\bibfnamefont{H.}~\bibnamefont{L{\"{u}}th}},
\bibinfo{journal}{Phys. Rev. B} \textbf{\bibinfo{volume}{55}},
\bibinfo{pages}{R1958} (\bibinfo{year}{1997}).








\bibitem[{\citenamefont{Mal'shukov and Chao}(2000)}]{Malshukov:prb00}
\bibinfo{author}{\bibfnamefont{A.~G.} \bibnamefont{Mal'shukov}}
\bibnamefont{and} \bibinfo{author}{\bibfnamefont{K.~A.}
\bibnamefont{Chao}}, \bibinfo{journal}{Phys. Rev. B}
\textbf{\bibinfo{volume}{61}}, \bibinfo{pages}{R2413}
(\bibinfo{year}{2000}).


\bibitem[{\citenamefont{Overhauser}(1953)}]{Overhauser:pr53}
\bibinfo{author}{\bibfnamefont}{A.~W.}~\bibnamefont{Overhauser},
\bibinfo{journal}{Phys. Rev.} \textbf{\bibinfo{volume}{92}},
\bibinfo{pages}{411} {\bibinfo{year}{1953}}

\bibitem[{\citenamefont{Salis}(2001)}]{Salis:prl01}
\bibinfo{author}{\bibfnamefont}{G.}~\bibnamefont{Salis},
\bibinfo{author}{\bibfnamefont}{D.~T.}~\bibnamefont{Fuchs},
\bibinfo{author}{\bibfnamefont}{J.~M.}~\bibnamefont{Kikkawa},
\bibinfo{author}{\bibfnamefont}{D.~D.}~\bibnamefont{Awshalom},
\bibinfo{author}{\bibfnamefont}{Y.}~\bibnamefont{Ohno},
\bibinfo{author}{\bibfnamefont}{H.}~\bibnamefont{Ohno},
\bibinfo{journal}{Phys. Rev. Lett.} \textbf{\bibinfo{volume}{86}},
\bibinfo{pages}{2677} {\bibinfo{year}{2001}}


\bibitem[{\citenamefont{Paget}(1977)}]{Paget:prb77}
\bibinfo{author}{\bibnamefont}{D.}~\bibnamefont{Paget},
\bibinfo{author}{\bibnamefont}{G.}~\bibnamefont{Lampel},
\bibinfo{author}{\bibnamefont}{B.}~\bibnamefont{Sapoval},
\bibinfo{author}{\bibnamefont}{V.~I.}~\bibnamefont{Safarov},
\bibinfo{journal}{Phys. Rev.} \textbf{\bibinfo{volume}{15}},
\bibinfo{pages}{5780} {\bibinfo{year}{1977}}

\bibitem[{\citenamefont{Berger}(1999)}]{Berger:prb99}
\bibinfo{author}{\bibfnamefont{L.}~\bibnamefont{Berger}},
\bibinfo{journal}{Phys. Rev. B} \textbf{\bibinfo{volume}{59}},
\bibinfo{pages}{11465} (\bibinfo{year}{1999});
\bibinfo{journal}{J. Appl. Phys.} \textbf{\bibinfo{volume}{90}},
\bibinfo{pages}{4632} (\bibinfo{year}{2001}).

 

\end{thebibliography}
\end{document}